\documentclass[conference]{IEEEtran}
\IEEEoverridecommandlockouts
\usepackage{cite}
\usepackage{amsmath,amssymb,amsfonts}
\usepackage{algorithmic}
\usepackage{graphicx}
\usepackage{textcomp}
\usepackage[margin={0.7in}]{geometry}
\def\BibTeX{{\rm B\kern-.05em{\sc i\kern-.025em b}\kern-.08em
    T\kern-.1667em\lower.7ex\hbox{E}\kern-.125emX}}
\begin{document}

\title{UAV Data Collection over NOMA Backscatter Networks: UAV Altitude and Trajectory Optimization
\thanks{This work was
supported by the European Union’s Horizon 2020 Research and Innovation Programme under Marie Sklodowska-Curie grant agreement no. 690893.}
}

\author{\IEEEauthorblockN{Amin Farajzadeh}
\IEEEauthorblockA{\textit{\rm Faculty of Engineering}\\ \textit{\rm and Natural Sciences} \\
\textit{\rm Sabanci University}\\
Istanbul, Turkey \\
aminfarajzadeh@sabanciuniv.edu}
 \and
\IEEEauthorblockN{Ozgur Ercetin}
\IEEEauthorblockA{\textit{\rm Faculty of Engineering}\\ \textit{\rm and Natural Sciences} \\
\textit{\rm Sabanci University}\\Istanbul, Turkey\\
oercetin@sabanciuniv.edu}
 \and
\IEEEauthorblockN{Halim Yanikomeroglu}
  \IEEEauthorblockA{\textit{\rm Department of Systems}\\ \textit{\rm and Computer Engineering} \\
\textit{\rm Carleton University}\\Ottawa, Canada\\
halim.yanikomeroglu@sce.carleton.ca}
}

\maketitle

\begin{abstract}
The recent evolution of ambient backscattering technology has the potential to provide long-range and low-power wireless communications. In this work, we study the unmanned aerial vehicle (UAV)-assisted backscatter networks where the UAV acts both as a mobile power transmitter and as an information collector. We aim to maximize the number of successfully decoded bits in the uplink while minimizing the UAV's flight time by optimizing its altitude. Power-domain NOMA scheme is employed in the uplink. An optimization framework is presented to identify the trade-off between numerous network parameters, such as UAV's altitude, number of backscatter devices, and backscatter coefficients. Numerical results show that an optimal altitude is computable for various network setups and that the impact of backscattering reflection coefficients on the maximum network throughput is significant. Based on this optimal altitude, we also show that an optimal trajectory plan is achievable.   
\end{abstract}

\begin{figure*}[!t]
\centerline{\includegraphics[width=152mm,scale=2]{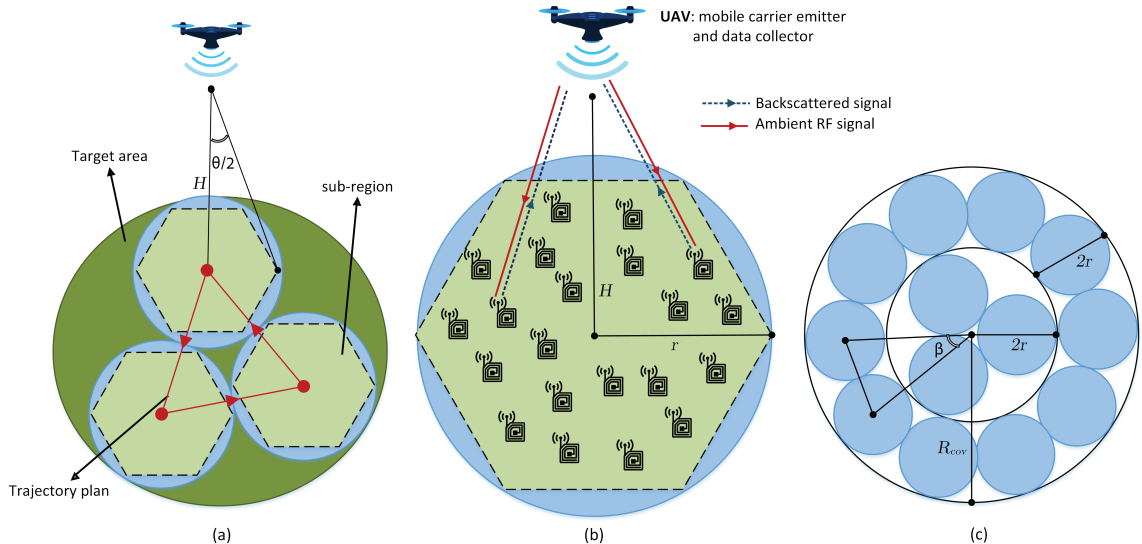}}
\caption{(a) Network model: Target area with hexagonal sub-regions and the trajectory plan, when the UAV is at an altitude $H$ with an effective illumination angle $\theta$. (b) Backscattering setup in one sub-region. (c) Geometry of dividing the target area into sub-regions (No of discs: $M=2$).} 
\label{fig1}
\end{figure*}
\section{Introduction}
\subsection{Motivation}
Ambient backscatter communication technology is a promising candidate for self-sustainable
wireless communication systems in which there is no external power supply \cite{b1a}. By utilizing the existing radio frequency (RF) signal, ambient
backscattering technology can support low-power sensor-type devices in the internet of things (IoT) paradigm \cite{b1}. In order to support a long-range backscatter communication link the following are needed: 1) A backscatter transmitter (tag), 2) a backscatter receiver (reader, data collector), and 3) one (or multiple) carrier emitter (RF energy source); it should be noted that the emitter may be collocated with the receiver \cite{ab133}. This novel technology allows to leverage the existing receiver
for generating the carrier signal. The state-of-the-art backscatter technology involves the design
of a novel backscatter tag that modulates the carrier
signal providing long-distance communication while consuming only {\textmu}Ws of power. For instance, the architectures proposed in \cite{b133} and \cite{b3} promise a long-range backscatter communication. Specifically, \cite{b3} achieves a range beyond
$3.4$ km when operating in the $868$ MHz band, and $225$ m when
operating in the $2.4$ GHz band which is a significant improvement
over the contemporary in backscatter communications. Hence, through the utilization of designs such as those described in \cite{b133} and \cite{b3}, wide-area communication is enabled by new passive backscatter IoT devices.\\
\indent Unmanned aerial vehicles (UAVs), also commonly known
as drones, have gained wide popularity in the recent
years for a variety of applications, such as cargo delivery and aerial imaging \cite{b1aa}.  
In particular, employing
UAVs as aerial base stations is envisioned as a promising solution
to improve the performance of the terrestrial wireless networks \cite{b2a}. Similarly, there has been a growing research interest in using UAVs for data collection and
dissemination in wireless networks, in order to provide a faster and reliable data collection, longer network
lifetime, and real-time data transmission \cite{bb2}, \cite{bbb2}. UAVs have great potential to be employed in long-range backscatter networks to both support more devices and increase the network efficiency and reliability. Consequently, optimizing the 3-D location of the data collecting UAV is very critical in order to provide reliable communication for backscatter devices which operate in the presence of very low power RF signals.
\\
\indent Recently, power-domain non-orthogonal multiple access (NOMA) is envisaged as an essential enabling technology for 5G wireless networks especially for uncoordinated transmissions \cite{b2b}. NOMA
exploits the difference in the channel gain among users for multiplexing. By allowing multiple users
to be served in the same resource block (to be decoded using successive interference cancellation (SIC)), NOMA may greatly improve the spectrum efficiency and may outperform traditional orthogonal multiple access schemes in many scenarios \cite{b2}. Moreover, it can support massive connectivity, since a large number of users can be served simultaneously \cite{b8a}. Motivated by this, in this paper, a network of long-range passive backscatter devices served by a UAV (used as both the emitter and data collector) are considered to access the medium based on the power-domain NOMA protocol.
\subsection{Related Works}
In the literature, there are many studies on optimizing the 3-D location of the aerial base stations under various scenarios. For instance, in \cite{b8}, the authors aim to optimize the UAV's altitude and antenna beamwidth for throughput maximization in three different communication models without considering the impact of altitude and beamwidth on the flight time. In \cite{b9}, a particle swarm optimization algorithm is proposed to find an efficient 3D placement of a UAV that minimizes the total transmit power required to cover the indoor users without discussing the outage performance and its dependency on the UAV's altitude. The impact of the altitude on the coverage range of UAVs was studied in \cite{b10}. In \cite{b11a}, an optimum placement of multiple UAVs for maximum number of covered users is investigated. In \cite{b11}, the authors aimed to find the optimal altitude which maximizes the reliability and coverage range. They consider the dependence of the path-loss exponent and multi-path fading on the height and angle of the UAV; however, similar to the previous works, they do not consider the impact of UAV's altitude on its flight time. Another drawback of the previous approaches is the lack of discussion on the control of ground networks with limited or no energy supplies. In this work, we consider passive devices which have no power supply, and investigate how their passive nature can impact the network performance.\\
\indent In addition, in \cite{bb2} and \cite{bbb2}, the authors consider a scenario where an UAV collects data from a set of sensors. In particular, in \cite{bb2}, they jointly optimize the scheduling policy and UAV's trajectory to minimize the maximum energy consumption of all sensors, while ensuring that the required amount of data is collected reliably from each node. In \cite{bbb2}, the authors investigate the flight time minimization problem for completing the data
collection mission in a one-dimensional sensor network. The objective is to minimize the UAV's total flight time from a starting point to a destination
while allowing each sensor to successfully upload a certain amount of data using a given amount of
energy. However, in these works, all the ground nodes are active devices which access the channel based on the conventional medium access control (MAC) protocols. \\
\indent In \cite{b12}, the authors investigate the applicability of NOMA for UAV-assisted communication systems. It is shown that the performance of NOMA scheme is far better than the orthogonal multiple access scheme under a number of different scenarios. Furthermore, in \cite{b13}, a NOMA-based terrestrial backscatter network is studied where the results suggest that NOMA has a good potential for being employed in backscatter communications.
\subsection{Contributions}
In this paper, we study the uplink of a UAV-assisted wireless network using power-domain NOMA where the ground nodes are backscatter devices. The main contributions of this paper are summarized as follows:
\begin{itemize}
\item We develop a framework where the UAV is used as a replacement to conventional terrestrial data collectors in order to increase the efficiency of collecting data from a field of passive backscatter nodes.
\item The objective is to maximize the number of successfully decoded bits while minimizing the flight time by determining the UAV's optimal altitude. Intuitively, if the UAV operates in lower altitudes, it would receive signals from fewer backscatter nodes, which would reduce interference in the NOMA setting; however, this operation increases the flight time since the UAV spends more time to cover the entire target area. On the other hand, when the UAV operates in higher altitudes, the quality of the received power is decreased due to excessive interference and high path-loss effect while the flight time is minimized. We show that there exists an optimal altitude at which the trade-off between the number of successfully decoded bits and the flight time duration is most favorable when the objective is to maximize the ratio of the number of successfully decoded bits to the flight time. Moreover, based on this optimal altitude, we show that an optimal trajectory plan is also achievable. 
\item In the MAC layer, instead of using a conventional orthogonal medium access schemes (e.g., time-division multiple access (TDMA)), we employ uplink power-domain NOMA scheme to
effectively serve a large number of passive backscatter nodes. 
\item The network throughput is defined as the ratio of the total number of successfully decoded bits to the whole flight time. Numerical results illustrate the optimal behavior of the UAV and the backscatter devices under
different scenarios. In particular, the dependency of the optimal altitude to various network parameters is analyzed which provides insight into the network behavior and design parameters. 
\end{itemize}                               
\subsection{Organization}
The rest of the paper is organized as follows. We describe the system model and background in Section II. Section III presents the problem formulation. The numerical results are discussed in Section IV. Finally, Section V concludes this paper.
\section{System Model and Background}
We consider a UAV-assisted backscatter network where $N$ backscatter nodes (BNs) are distributed independently and uniformly (i.e., binomial point process) with a sufficiently high density in a target area on the ground and there is a single UAV acting as both mobile power transmitter and information collector. The network model is illustrated in Fig.~\ref{fig1}.a. We assume that the UAV is equipped with a directional antenna
with fixed effective illumination angle (or beamwidth) and it hovers over the target area for a duration of $T_{f}$. During the total flight time $T_{f}$, the UAV continuously broadcasts a single carrier RF signal with fixed power $P_u$ to all BNs on the ground, i.e., it acts as carrier emitter. On the ground side, the BNs become active and utilize the received RF signal to backscatter their data to the UAV simultaneously based on power-domain NOMA scheme. 
\subsection{Channel Model}
In this work, we consider a path-loss model in which the channel power gain of the link between the UAV and BN $i$, $i=1,\dots,N$, is defined as $h_{BN_i}d_{BN_i}^{-\alpha}$ where $h_{BN_i}=10^{\frac{g_{BN_i}}{10}}$ denotes the shadowing effect following a log-normal distribution. $g_{BN_i}$ is a Normal distributed random variable, with zero mean and standard deviation, $\sigma$, which is typically between $0$ and $10$ dB. Moreover, $d_{BN_i}^{-\alpha}$ denotes the distance-dependent attenuation in which $\alpha$ is the path-loss exponent and $d_{BN_i}$ is the distance between BN $i$ and the UAV.
 In the following, we provide a brief overview of the ambient backscattering and power domain NOMA as employed in this paper.
\subsection{Ambient Backscattering}
Upon receiving RF signal from the UAV, the BNs use a modulation scheme, such as FSK, to map their data bits to the received RF signal and then backscatter them to the UAV, simultaneously, for a duration of $T$ time units. After the transmission, BN switches to the sleep mode and remains at this mode until the end of the UAV's total flight time.
The received power at BN $i$ can be written as
\begin{align}
P_{BN_i}^{rx}=P_uh_{BN_i}d_{BN_i}^{-\alpha}.
\end{align}
Let $\zeta_{BN_i}$ be the reflection coefficient of BN $i$. Thus, the power of the backscattered signal at each BN is determined as,
\begin{align}
P_{BN_i}^{tx}=\zeta_{BN_i} P_{BN_i}^{rx}.
\end{align}
Moreover, we assume that the data rate for each BN is $R$ bits$/$secs which is a constant since the rate is controlled by the setting of the circuit elements in backscatter devices \cite{b20}.  
\subsection{NOMA Protocol}
In this work, we consider a power-domain NOMA scheme as the uplink MAC protocol. In order for NOMA scheme to be able to successfully decode the incoming signals, the difference of the channel gains on the same spectrum resource must be sufficiently large \cite{b13}. Thus, it is assumed that the channel power gains of BNs in each sub-region, which is discussed in Sec. II.D, are distinct and can be ordered which is a common assumption in the uplink NOMA scenario \cite{b20b}\cite{b13b}. Under this assumption, the product of uplink and downlink channel gains can be ordered as
\begin{align}\label{3}
d_{k_1}^{-2\alpha}h_{k_1}^2>\dots>d_{k_{N_l}}^{-2\alpha}h_{k_{N_l}}^2,
\end{align} 
where $k_{(.)}\in \{BN_1,\dots,BN_N\}$ such that $k_1,\dots,k_{N_l}$ represent the BNs in sub-region $s_l$, $l=1,\dots,W$, and $N_l$ is the number of BNs in sub-region $s_l$ such that $N=\sum_{l=1}^W N_l$. Moreover, to make the difference of channel gains more pronounced and obtain a diverse set of received powers, all BNs at each sub-region backscatter their data to the UAV simultaneously with different reflection coefficients,
\begin{align}\label{gamma}
1>\zeta_{k_1}>\dots>\zeta_{k_{N_l}}>0,
\end{align}
such that with SIC employed at the UAV, the successful retrieval and decoding of the BNs' signals become possible. In order to assign reflection coefficients to BNs, the following approach is adopted at the UAV: Since the UAV knows the exact location of BNs, it is accordingly aware of the distances from them at any given altitude and sub-region. Hence, assuming that each BN has a unique ID, the UAV assigns the highest reflection coefficient to the closest BN and, in a descending order, assigns the lowest reflection coefficient to the farthest BN at each sub-region. Note that we assume the time for assigning reflection coefficients is negligible compared to the backscattering time $T$.\\  
\indent The best performance of NOMA scheme is achieved when the signal-to-interference-plus-noise ratio (\textrm{SINR}) for each one of the backscattered signals at the UAV is greater than a given SINR threshold $\gamma$ necessary for successful decoding. This implies the following: 
\begin{align}\label{sinr}
\textsf{SINR}_{BN_i}=\frac{P_u\zeta_{BN_i}h_{BN_i}^2d_{BN_i}^{-2\alpha}}{\sum_{j=i+1}^{N_l}P_u\zeta_{BN_j}h_{BN_j}^2d_{BN_j}^{-2\alpha}+\mathcal{N}}\geq \gamma,\\
\forall\: i=1,\dots,N_l, \hspace{2.5cm}\nonumber
\end{align}
where $\mathcal{N}$ is the noise power. Note that we assume that the backscattered signal by $k_1$ is the strongest signal at each sub-region and gets decoded at the UAV first; on the other hand, $k_{N_l}$'s signal is considered to be the weakest one and gets decoded after all the stronger signals are decoded \cite{b13}. 
\subsection{UAV's Mobility Model}
We assume that the coverage area of the UAV when it flies at an altitude $H_{max}$ with an effective illumination angle $\theta$ is a circle with radius $R_{cov}=H_{max}\tan(\frac{\theta}{2})$. This circle covers the whole target area that is assumed to be in a hexagonal shape with diameter $2R_{cov}$. In order to improve the number of successfully decoded bits, the UAV may need to lower its altitude to get closer to BNs, and thus, it cannot cover the entire target area in a single time slot; in this case, the target area is divided into $W$ sub-regions each with the same radius such that at an altitude of $H$, the sub-region radius is determined as $r=H\tan(\frac{\theta}{2})$. Consequently, the total flight time will be divided into $W$ sub-slots (ignoring the time it takes to fly from one sub-region to the other). To determine the number of sub-regions (equivalently, sub-slots) needed to cover the entire target area, we first divide the target area covered at altitude $H_{max}$ into $M$ discs with the same center and radius difference of $2r$, which is obtained as 
\begin{align}
M=\begin{cases}
 \left\lfloor \frac{R_{cov}}{2r} \right\rfloor,  \text{ }\text{if } r\leq\frac{R_{cov}}{2},\\
1,\quad \qquad\text{if } r>\frac{R_{cov}}{2},
\end{cases}
\end{align}
where $\left\lfloor x \right\rfloor$ is the floor function mapping $x$ to the greatest integer value less than or equal to $x$. Then, the number of sub-regions with radius $r$ inside disc $m$, where $m=1,\dots,M$, is calculated by
\begin{align}
w_m=\left\lfloor \frac{2\pi}{\beta_m}\right\rfloor,
\end{align}
where $\beta_m$ is the angle between two adjacent sub-regions with respect to the origin point such that $\sin(\frac{\beta_m}{2})=\frac{r}{R_{cov}-(2m-1)r}$. Hence, the total number of sub-regions covered by the UAV can be determined as
\begin{align}
W=
\begin{cases}
\sum_{m=1}^{M}w_m, & \text{if } H_{min}\leq H<H_{max},\\
1,& \text{if } H=H_{max}.
\end{cases}
\end{align}
The number of sub-regions $W$ implies that the UAV's total flight time, $T_{f}$, is divided into $W$ sub-slots with the same duration of $T$ assuming that the UAV's flying speed is sufficiently high \cite{b21}, i.e.,
\begin{align}\label{9}
T_{f}\approx WT.
\end{align}
When $W=1$, it means that there is no sub-region and the UAV remains at altitude $H=H_{max}$ during $T_f=T$. Fig.~\ref{fig1}.c illustrates the geometry of dividing the target area into sub-region. Note that the BNs are served by the UAV only once since each BN switches to sleep mode until the end of UAV's flight time after backscattering its data. \\
\indent Let $(x, y, H)$ be the 3-D coordinate of UAV. Thus, the distances between the UAV and any BN can be calculated as
\begin{align}
d_{BN_i}=\sqrt{H^2+(x_{BN_i}-x)^2+(y_{BN_i}-y)^2},
\end{align}
where $x_{BN_i}$ and $y_{BN_i}$ are the coordinates of BN $i$. In this work, we assume that the UAV knows the exact location of the BNs. Furthermore, the UAV's trajectory plan is modeled as: Given the number of sub-regions $W$ which is obtained at any altitude as discussed above, the UAV moves from the origin of each sub-region as its 2-D location over each sub-region, i.e., $(x,y)$, to adjacent sub-region as illustrated in Fig.~\ref{fig1}.a. According to \eqref{9}, since we assume that the flying time from each origin to adjacent one is negligible compared to the flight time over each sub-region, it does not matter the UAV starts to hover from which sub-region first. \\ 
\begin{table*}[!t]
\caption{Simulation Altitudes}
\begin{center}
\begin{tabular}{|c|c|c|c|c|c|c|c|c|c|c|}
\hline
Altitudes $H$ ($\textrm{m}$) & $86.71$& 80.71 & 72.21&64.21&58.21&52.71& 48.21& 44.21& 43.71& 43.21 \\
\hline
Number of sub-regions $W$ &1 & 2& 3 & 4& 5&6&7&8&9&12\\
\hline
Number of BNs at each sub-region $N_l$&40&20&13&10&8&7&6&5&4&3\\
\hline
\end{tabular}
\label{tab2}
\end{center}
\end{table*}
\begin{table}[!t]
\caption{Simulation Parameters}
\begin{center}
\begin{tabular}{|c|c|}
\hline
\textbf{Parameter} & \textbf{\textbf{Value}}
\\
\hline
Total number of BNs ($N$) & $40$  \\
\hline
Effective
illumination angle ($\theta$) & $60^\circ$ \\
\hline
UAV transmit power ($P_u$) \cite{b3} & $20 \textrm{ dBm}$\\
\hline
Noise power ($\mathcal{N}$) & $-70 \textrm{ dBm}$\\
\hline
Transmission rate ($R$) & $64\textrm{ bits/sec}$\\
\hline
Radius of target
area ($R_{cov}$) & $100\textrm{ m}$\\
\hline
SINR threshold ($\gamma$) & $-3$ \textrm{ dB}\\
\hline
Path-loss exponent ($\alpha$) & $2.7$\\
\hline
Reflection coefficient range ($\zeta$) & $[0.1,0.99]$\\
\hline
Maximum number of sub-regions ($W_{max}$) & $12$\\
\hline
Log-normal shadowing variance ($\sigma^2$) & $8$ \textup{ dB}\\
 \hline
\end{tabular}
\label{tab1}
\end{center}
\end{table}
\section{Problem Formulation}
Our objective is to maximize the total number of successfully decoded bits by the UAV while minimizing its flight time, by finding an optimal altitude $H^*$. We consider an application scenario, where data from {\it all} BNs within the sub-region should be successfully decoded. Otherwise, the whole sub-region data is discarded. This metric is appropriate when each BN's data is unique and uncorrelated, and thus, it is a requirement to collect data from all BNs. Hence, we define the network throughput $C(H)$ as the ratio of the total number of successfully decoded bits during all time sub-slots (i.e., in all sub-regions) to the total flight time:
\begin{align}\label{throughput}
C(H)=\frac{\sum_{l=1}^{W(H)} C_l(H)}{T_f(H)},
\end{align}
where
\begin{align}\label{succ_bits}
C_l(H)=N_l(H)TR(1-
P_{out}^{(s_l)}(H)),
\end{align}
is the number of successfully decoded bits at sub-region $s_l$, $l=1,\dots,W$, and also $P_{out}^{(s_l)}(H)$ is the outage probability corresponding to sub-region $s_l$, which is determined as\footnote{In order to simplify the notation, from now on we will not show the $H$ dependence explicitly; for instance, we will use $C$ instead of $C(H)$.}
\begin{align}\label{12}
P_{out}^{(s_l)}
=1-\textrm{Pr}(\textsf{SINR}_{{k_1}}^{(s_l)}\geq \gamma, \dots, \textsf{SINR}_{{k_{N_l}}}^{(s_l)}\geq \gamma).
\end{align}
By using \eqref{3}, \eqref{gamma} and \eqref{sinr}, we have
\begin{align}
P_u\zeta_{k_1}h_{k_1}^2d_{k_1}^{-2\alpha} &\geq P_u\zeta_{k_2}h_{k_2}^2d_{k_2}^{-2\alpha}\gamma\nonumber\\&+\gamma\sum_{ j=3}^{N_l}P_u\zeta_{k_j}h_{k_j}^2d_{k_j}^{-2\alpha}+\gamma\mathcal{N}\nonumber \\ &\approx   \gamma\sum_{ j=3}^{N_l}P_u\zeta_{k_j}h_{k_j}^2d_{k_j}^{-2\alpha}+\gamma\mathcal{N}.
\end{align}
This approximation holds due to the distinct channels gains and reflection coefficients which are stated in \eqref{gamma} and \eqref{sinr}, respectively. Consequently, $P_u\zeta_{k_1}h_{k_1}^2d_{k_1}^{-2\alpha}\gg P_u\zeta_{k_2}h_{k_2}^2d_{k_2}^{-2\alpha}\gamma$ assuming $\gamma\leq 1$, and thus, $P_u\zeta_{k_2}h_{k_2}^2d_{k_2}^{-2\alpha}$ has infinitesimal effect on $\textrm{Pr}(\textsf{SINR}_{k_1}\geq \gamma)$ compared to $\gamma\sum_{ j=3}^{N_l}P_u\zeta_{k_j}h_{k_j}^2d_{k_j}^{-2\alpha}$ \cite{b22}. Hence, the events $\textsf{SINR}_{k_1}\geq \gamma$ and $\textsf{SINR}_{k_2}\geq \gamma$ are approximately independent. The same argument can be applied
to argue that $\text{Pr}(\textsf{SINR}_{k_i}\geq \gamma|\textsf{SINR}_{k_{i^\prime} }\geq \gamma)\approx \text{Pr}(\textsf{SINR}_{k_i}\geq \gamma)$ for any $i<i^\prime$ where $i\geq 2$. Therefore, \eqref{12} can be approximated as
\vspace{-0.15cm}
\begin{align}
P_{out}^{(s_l)}
\approx 1-\prod_{j=1}^{N_l} \textrm{Pr}(\textsf{SINR}_{k_j}^{(s_l)}\geq \gamma).
\end{align}
Define $z_i=\zeta_{k_i}h_{k_i}^2d_{k_i}^{-2\alpha}$, $i=1,\dots,N_l$, which is a log-normal distributed random variable since the product of two log-normal distributed random variables is also log-normal with mean $\mu_{z_i}=\ln(\zeta_{k_i}d_{k_i}^{-2\alpha})$ and variance $\sigma_{z_i}^{2}=4a^2 \sigma^{2}$ where $a=\frac{\ln 10}{10}$. Then, we have (from \eqref{sinr})
\begin{align} \label{16}
\textrm{Pr}(\textsf{SINR}_{k_i}\geq \gamma)=\textrm{Pr}(\frac{z_i}{\sum_{j=i+1}^{N_l} z_j+\frac{\mathcal{N}}{P_u}}\geq\gamma).
\end{align}
To make the problem tractable, we assume that the thermal noise is negligible and it is only taken into account when there is no interference (i.e., in calculating the SINR of the weakest BN at each sub-region $\textsf{SINR}_{k_{N_l}}$) \cite{b23a}. The distribution of $\sum_{j=i+1}^{N_l} z_j$ has no closed-form expression, but it can be reasonably approximated by another log-normal distribution $A_i$ at the right tail. Its probability density function at the neighborhood of $0$ does not resemble any log-normal distribution \cite{b23} \cite{b23b}. Using the Fenton-Wilkinson method \cite{b24}, a commonly used approximation is obtained by matching the mean and variance of another log-normal distribution as
\begin{align}
&\mu_{A_i}=\ln \left[\sum_{j=i+1}^{N_l} e^{\mu_{z_j}+\frac{\sigma^2_{z_j}}{2}}\right]-\frac{a^2\sigma^2_{A_i}}{2},\hspace{1.8cm} \\
&\sigma_{A_i}^2=\ln\left[\frac{\sum_{j=i+1}^{N_l} e^{(2\mu_{z_j}+\sigma^2_{z_j})}(e^{\sigma^2_{z_j}}-1)}{(\sum_{j=i+1}^{N_l} e^{\mu_{z_j}+\frac{\sigma^2_{z_j}}{2}})^2}+1\right].
\end{align}
Thus, $\textsf{SINR}_{BN_{(.)}}$ can be approximated by a log-normal random variable defined as $Y_{BN_{(.)}}$ with mean $\mu_{Y_{(.)}}$ and variance $\sigma_{Y_{(.)}}^2$, which can be calculated as 
\begin{align}
\mu_{Y_i}=
\begin{cases}
\mu_{z_i}-\mu_{A_i},\:& \forall\: i\neq N_l,\\
\mu_{z_i}-\ln(\frac{\mathcal{N}}{P_u}),\:&\forall\: i=N_l,
\end{cases}
\end{align}
and
\begin{align}
\sigma_{Y_i}^2=
\begin{cases}
\sigma_{z_i}^2+a^2\sigma_{A_i}^2,\: &\forall\: i\neq N_l,\\
\sigma_{z_i}^2,\:&\forall\: i=N_l.
\end{cases}
\end{align}Hence, the outage probability corresponding to sub-region $s_l$ can be determined as
\begin{align}
P_{out}^{(s_l)}&\approx 1-\prod_{j=1}^{N_l} \textrm{Pr}(Y_{k_j}\geq \gamma)\nonumber \\
&=1-\prod_{j=1}^{N_l}\left[\frac{1}{2}-\frac{1}{2}\textrm{erf}(\frac{10\log_{10} (\gamma )-\mu_{Y_{j}}}{\sigma_{Y_{j}}\sqrt{2}})\right].
\end{align}
Finally, the optimization problem can be expressed as 
\begin{align}\label{21}
&\operatorname*{max}_{H\in\mathcal{H}} C \nonumber\\
&\text{s.t.}\:
1\leq W\leq W_{max},
\end{align}
where $\mathcal{H}\in \{ H_{min},\dots, H_{max}\}$ is a set of discrete altitudes. Note that $H_{max}$ corresponds to $W=1$, and $H_{min}$ to $W=W_{max}$. The set of altitudes is determined by the operational requirements of the UAV.
Furthermore, \eqref{21} is
a fractional programming (FP) problem with non-differentiable fractional objective function. Since the cardinality of the set of altitudes that a UAV can hover over is finite, and the locations of BNs are known a priori, we use exhaustive search to determine the optimal solution. 
\begin{figure}[!t]
\centerline{\includegraphics[width=99mm,scale=10]{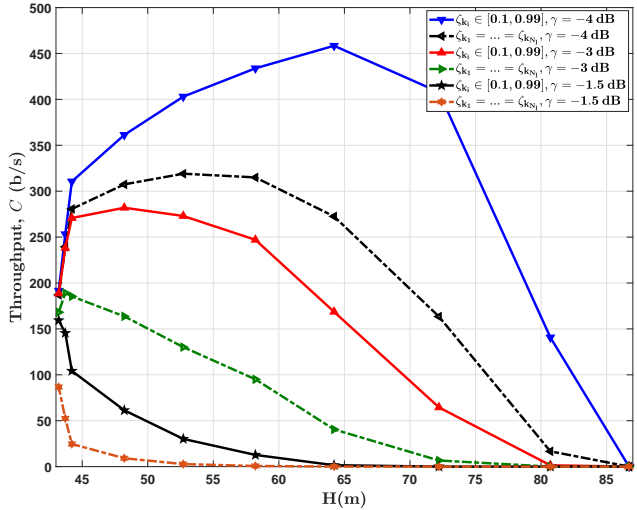}}
\caption{Throughput performance with respect to UAV altitude $H$, for two different ways of selecting the selection of reflection coefficients $\zeta$ and for three different SINR thresholds $\gamma$ ($N=40,\alpha=2.7$).}
\label{fig2}
\end{figure}
\begin{figure}[!t]
\centerline{\includegraphics[width=99mm,scale=10]{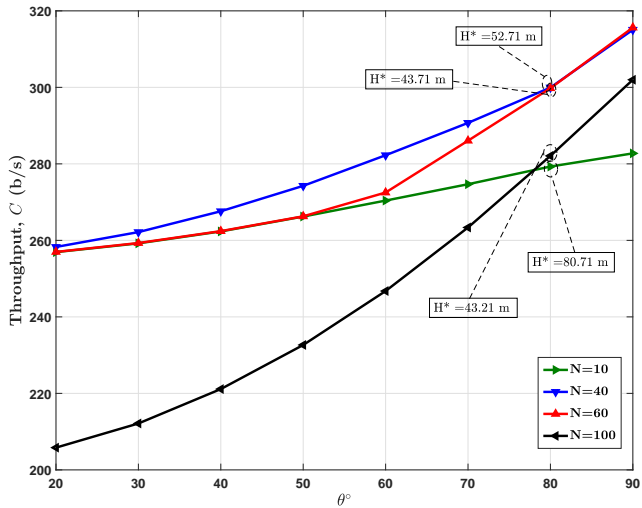}}
\caption{Throughput performance at optimized altitude 
with respect to the effective illumination angle $\theta$ for different considerations for the total number of BNs $N$ ($\gamma=-3$ \textmd{dB}, $\alpha=2.7$).}
\label{fig3}
\end{figure}
\begin{figure}[!t]
 \centerline{\includegraphics[width=99mm,scale=10]{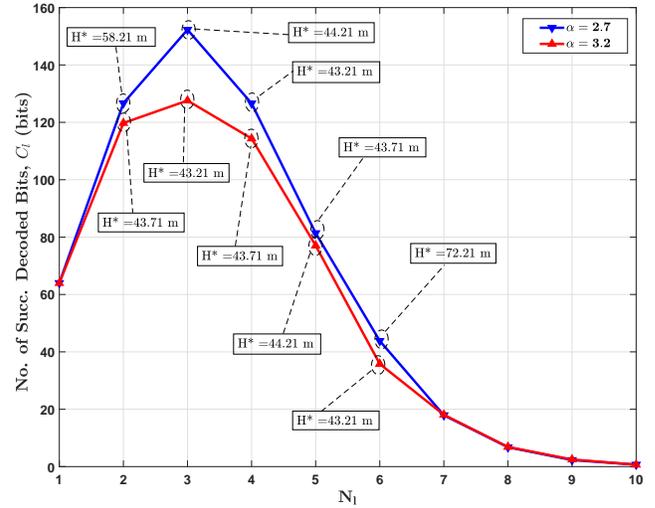}}
\caption{The performance of the number of successfully decoded bits (Eq. \eqref{succ_bits}) corresponding to $s_l$ at optimized altitude with respect to the number of BNs $N_l$ for different path-loss exponents $\alpha$ ($\gamma=-1.5$ \textmd{dB}).}
\label{fig4}
\end{figure}
\section{Numerical Results}
In this section, we evaluate the throughput $C$ with respect to UAV's altitude, under various considerations of network parameters including the SINR threshold $\gamma$ and backscattering reflection coefficients. We also analyze the effect of the effective illumination angle $\theta$ with different consideration for the total number of BNs $N$, on the throughput at optimized altitude. Moreover, the dependency of the number of successfully decoded bits $C_l$ at sub-region $s_l$ on the number of BNs $N_l$ inside the sub-region is investigated for two different path-loss exponents $\alpha$. A discrete set of UAV altitude is given in Table \ref{tab2} calculated using the procedure outlined in Sec. II.D with a target area radius of $100$ m. Unless otherwise stated, in all experiments we use the parameters given in Table \ref{tab1}.\\
\indent In Fig.~\ref{fig2}, the throughput is plotted with respect to $H$ for $\gamma=-4$, $-3$, and $-1.5$ \textrm{dB}. The figure illustrates that with lower SINR thresholds, there exists an optimal altitude where the throughput is maximized, and as the sensitivity of the SIC decoder at UAV increases, the throughput increases as well. As the altitude is high, the number of BNs backscattering is also high, but the received power from each are close. This reduces the probability of correct decoding. However, if the altitude is low, then even if there are fewer incoming transmissions from the BNs, the total flight time of the UAV is high, reducing the throughput. In Fig.~\ref{fig2}, we also examine the performance of the network throughput with respect to UAV’s altitude $H$ with different BN reflection coefficients. The figure shows that the way the reflection coefficients are selected has a significant impact on the throughput (the network parameters used for Fig.~\ref{fig2} are given in Table II). When the reflection coefficients assigned to the $40$ BNs are in the range $[0.1, 0.99]$ with equal intervals at each sub-region (i.e., $\zeta_{k_{N_l}} = 0.1, \zeta_{k_{N_l-1}} =0.1 + \frac{(0.99-0.1)}{N_l-1}, \zeta_{k_{N_l-2}} = 0.1 + \frac{2(0.99-0.1)}{N_l-1}, ..., \zeta_{k_1} = 0.99$), the throughput improves by more than $40\%$ compared to the case when all the reflection coefficients are the same, for $\gamma = -4$ dB. When the reflection coefficient values are apart from each other, the received powers of the backscattered signals get further apart, and thus, the SIC decoder makes fewer decoding errors. Note that when $\zeta_{k_1} =\dots=\zeta_{k_{N_l}}$, the actual values of $\zeta_{k_{(.)}}$ does not matter due to the fact that, when the background noise is omitted in (16), the $\zeta_{k_{(.)}}$ values in the numerator and denominator will cancel each other.\\
\indent Furthermore, in Fig.~\ref{fig3}, we evaluate the performance of the throughput value at the optimized altitude with respect to the effective illumination angle $\theta$, under different considerations for the total number of BNs $N=10$, $40$, $60$, and $100$. The figure shows that the throughput at the optimized altitude monotonically increases as $\theta$ grows. When $\theta$ value is low, the UAV operates at an higher altitude to cover the target area, hence, the path-loss effect is notably high reducing the throughput. However, in high $\theta$ values, the UAV operates at a lower altitude. Thus, the throughput increases dramatically due to significant reduction in path-loss effect. Moreover, it can be seen that as $N$ increases from $10$ to $40$, the throughout improves which is due to the increase in the number of decoded bits. However, more increase of $N$, results in the domination of interference decreasing the throughput. When $\theta$ is above $80^\circ$, we also notice that the throughput, when $N=10$, is less than that of when $N=100$. This is because with these high $\theta$ values, the UAV operates at lower altitudes where the path-loss effect is low.\\
\indent Finally, in Fig.~\ref{fig4}, we investigate the dependency of the number of successfully decoded bits at optimized altitude at one sub-region to the number of BNs inside the sub-region for different path-loss exponent values, $\alpha=2.7$ and $\alpha=3.2$. The figure shows that when the number of BNs is high at each sub-region, the outage probability increases due to the high interference. On the other hand, when this number is low, even though the decoding outage is low, fewer number of bits get decoded; hence, the curve decreases dramatically. Moreover, the figure implies that for each environment, there exist a pair of optimal altitude and number of BNs, i.e., ($H^*,N_l^*$), such that the number of successfully decoded bits at one sub-region is maximized. It can also be seen that as the environment gets more lossy, this number decreases dramatically by more than $19\%$ around the peak value.
\section{Conclusion}
In this paper, we studied the performance of a novel network model where a NOMA-based long-range backscatter network is facilitated with an aerial power station and data collector. Our objective was to investigate the relationship between the optimal altitude of the UAV and the total number of successfully decoded bits and the UAV's flight time. To the best of the author's knowledge, this is the first work in the literature which studies the UAV-enabled backscatter networks where the objective is to maximize the number of successfully decoded bits while minimizing the flight time by finding the UAV's optimal altitude. The results show that for a selection of parameters, there exist an optimal altitude where the ratio of the number of successfully decoded bits to the flight time is maximized. In the future work, we will consider more realistic scenarios where the UAV does not know the exact location of BNs. Hence, the problem will involve stochastic geometry which will add more complexity to the problem. Moreover, the design framework
can also be extended to the multi-UAV scenario, where
the UAV-BN association and co-channel interference should
be taken into account. 

\end{document}